\documentclass[]{article}
\newcommand{\bb}{\begin{eqnarray}}
\newcommand{\ee}{\end{eqnarray}}
\begin{document}
\title{Absence of log correction in entropy of large black holes}
\author{A. Ghosh\thanks{amit.ghosh@saha.ac.in}\\
P. Mitra\thanks{parthasarathi.mitra@saha.ac.in}\\
Saha Institute of Nuclear Physics\\ 1/AF Bidhannagar\\
Calcutta 700064}
\date{}
\maketitle
\begin{abstract}
Earlier calculations of black hole entropy in loop quantum gravity led to a
dominant term proportional to the area, but there was a correction
involving the logarithm of the area, 
the Chern-Simons level being assumed to be large.
We find that the calculations yield an entropy proportional to the area
eigenvalue with no such correction if the Chern-Simons level is finite, 
so that the area eigenvalue can be relatively large.
\end{abstract}

\section*{Introduction}
Loop quantum gravity has given a way of understanding black hole entropy
from a quantization of black hole spacetimes. Both the U(1) approach
\cite{ash} and the SU(2) approach \cite{perez} have yielded an area
dependence for the entropy. However, usually the entropy has, apart from a
linear term in the horizon area, a logarithmic correction term.
The U(1) formulation gives a coefficient $-1/2$ \cite{bhe1,bhe2,cor,sahlmann}
for the logarithmic term. The SU(2) formulation \cite{perez} gives a
coefficient $-3/2$ instead. This version of the theory was recently
elaborated upon for a finite Chern-Simons level $k$ \cite{k} and 
found to lead to an entropy with no logarithmic term for 
large area eigenvalues \cite{pm}
while a logarithmic term does appear for large area {\it and} large $k$. 
This is of significance because the Bekenstein-Hawking expression for the
entropy too is simply linear in the area with no correction.
In the present work we study similarly the original U(1) formulation for finite 
$k$. It has hitherto been considered only for large
$k$, where a logarithmic correction
term is present. We show that this term goes away for finite $k$.

The quantum theory of isolated horizons envisages punctures on the horizon as
some effective degrees of freedom, with spins associated with the punctures.
The number of punctures and the spins can vary. A puncture carrying spin-$j$
contributes an amount $8\pi\gamma\ell_P^2\sqrt{j(j+1)}$ to the quantum area,
where $\gamma$ is the so-called Immirzi parameter, which is a quantization 
ambiguity and $\ell_P$ is the Planck length.
One has to count the number of spin states which satisfy the condition that
{\em the sum of spin projections on the punctures is zero modulo}
$n=\frac12 k$, where $k$ is the level of the U(1) Chern-Simons theory
which is related to the classical horizon area involved in the formulation 
\cite{ash}. This
condition has mostly been simplified by requiring that the sum be zero,
which amounts to taking $k$ to be large. Even the numerical
counting of states in \cite{cor} used this simplifed projection constraint.


\section*{Spin 1/2}

Let us consider first the simplified case where all punctures
have spin one-half associated with them and let their number be an
even integer $p$. The total number of possible spin states is
\bb
2^p=\sum_{r=0}^p{}^pC_r,
\ee
where $r$ represents the number of punctures with spin up
and ${}^pC_r\equiv{p!\over r!(p-r)!}$, the binomial coefficient.
Let us take $p$ to be larger than $k=2n$ (where for simplicity, 
$k$ is taken to be an even integer) and furthermore, define $\ell$ by
\bb
\frac{p}{2}=\ell ~{\rm mod}~ n,\quad 0\leq\ell\leq n-1.
\ee
For zero spin projection, $r=p/2$, but for spin projection zero modulo $n$,
$r-p/2=0$ mod $n$, {\it i.e.,} $r=\ell$ mod $n$.
So the number of such states is simply
\bb
N={}^pC_{\ell}+{}^pC_{\ell+n}+...+{}^pC_{p-\ell-n}+{}^pC_{p-\ell}.
\ee
Now, for any $s$ between 0 and $n-1$, one has the binomial sum
\bb
(1+e^{2is\pi/n})^p=\sum_{r=0}^pe^{2irs\pi/n}~{}^pC_r.
\ee
Hence,
\bb
e^{-2i\ell
s\pi/n}(1+e^{2is\pi/n})^p=\sum_{r=0}^pe^{2i(r-\ell)s\pi/n}\;{}^pC_r.
\ee
Adding these equations for all values of $s$, one notices that on the right
hand side $\sum_se^{2i(r-\ell)s\pi/n}=n$ if
$r=\ell$ modulo $n$, while for other $r$, the sum vanishes. 
The binomial coefficients that survive are precisely the ones in $N$. Hence,
\bb
\sum_{s=0}^{n-1}e^{-2i\ell\pi s/n}(1+e^{2is\pi/n})^p=nN,
\ee
which is exact. If $p\gg n$, the sum
is dominated by the term of highest magnitude. Noticing that
\bb
1+e^{2is\pi/n}=2\cos (s\pi/n)e^{is\pi/n},
\ee
one sees that the highest magnitude occurs for $s=0$ and
\bb
N\approx 2^p/n,
\ee
the other terms being exponentially suppressed for large $p$ in the sense that
\bb
2^p+(2\cos\theta)^p=2^p[1\pm\exp(-p\ln|\sec\theta|)].
\ee
Now the area eigenvalue in the present situation is
$4\pi\gamma\ell_P^2\sqrt{3}p\equiv A$, so that the entropy is
\bb
\log N= A{\log 2\over 4\pi\gamma\ell_P^2\sqrt{3}}-\log n.
\ee
If the constant term is neglected, the entropy is proportional to $A$ only
and the sub-leading logarithmic correction is absent.

In earlier calculations, the number $n$ was taken to be large and
the simplified projection constraint with zero projection was imposed:
for $r$, only $\frac{p}{2}$ was considered instead of the
series of values $r=\ell, \ell+n,...p-\ell$ above. Using
Stirling's approximation 
\bb
n!\approx n^n\sqrt{2\pi n}\;e^{-n},\label{stirling}
\ee
one then obtained
\bb
{}^pC_{p/2}={p!\over (p/2)! (p/2)!}\approx {2^p\over\sqrt{p}},
\ee
so that the $\sqrt{A}$ coming from the denominator yielded a $-\frac12\log{A}$
contribution to the entropy \cite{bhe1,bhe2}.

\section*{The general spin case}

Next we consider the case of arbitrary spins associated with the punctures.
Let $s_{jm}$ be the number of punctures with spin quantum numbers $j,m$
in a certain configuration.
Then the number of all spin states is
\bb
\sum_{\{ s_{jm}\} }{(\sum_{jm} s_{jm})!\over \prod_{jm} (s_{jm}!)}.
\ee
Not all of these are allowed, because the spin projection condition has to
be imposed:
\bb
\sum_{jm}ms_{jm}=0,\label{spinp}
\ee
where the strict equality is imposed to begin with. Until the other
possibilities modulo $n$ are taken into account, this calculation
follows \cite{bhe2}. Only states with a definite area $A$ are considered:
\bb
\sum_{jm}8\pi\gamma\ell_P^2\sqrt{j(j+1)}s_{jm}=A.
\ee
To maximize the probability of a configuration $\{s\}$, one must have
the combinatorial factor for $\{s\}$ or its logarithm stationary, {\it i.e.,}
\bb
(\sum \delta s)\ln\sum s &-&\sum (\delta s\ln s)=0,
\ee
where the simplified version of Stirling's approximation has been used,
{\it i.e.,} the square root factor in (\ref{stirling}) ignored. This relation is
subject to
\bb
\sum m \delta s=0 
\ee
and
\bb
\sum\sqrt{j(j+1)}\delta s=0.
\ee
With two Lagrange multipliers to take the two constraints into account, we find
\bb
{s_{jm}\over \sum s}=e^{-2\lambda\sqrt{j(j+1)}-\alpha m}.
\ee
It follows that
\bb
1=\sum_{jm}e^{-2\lambda\sqrt{j(j+1)}-\alpha m}.
\label{cond}\ee
Up to this point the calculations are the same even if the vanishing total spin
projection is replaced by a fixed nonzero value.
The condition of vanishing spin projection sum implies that
\bb
\alpha=0,
\ee
though later we shall need a non-vanishing value of $\alpha$. The
combinatorial factor for $\{ s\}$ is then seen to reduce to
\bb
\exp({\lambda A\over 4\pi\gamma\ell_P^2})
\ee
in the simplified Stirling approximation, and there is an extra factor
\bb
{\sqrt{2\pi\sum\bar s}\over\prod\sqrt{2\pi\bar s}}
\label{sqrt}\ee
in the full Stirling approximation, 
where ${\bar s_{jm}}$ is the most probable configuration.
To take care of this piece, it is necessary
to expand $s_{jm}$ about ${\bar s_{jm}}$
and sum over the fluctuations. Because of the stationary
behaviour about the most probable configuration, the first order variation
vanishes and only the second order variations are retained in a gaussian
approximation. Thus one has
\bb
{(\sum s)!\over \prod s!}={(\sum \bar s)!\over \prod \bar s!}
\exp[-\sum{(\delta s)^2\over 2\bar s}+ {(\delta\sum s)^2\over 2\sum\bar s}].
\ee
If the second term in the exponent were absent,
each $\delta s=s-\bar s$ would produce on integration a factor equal to
$\sqrt{2\pi\bar s}$,
to be compared to a similar factor in the denominator of (\ref{sqrt}).
It may be noted that the second term in the exponent produces a zero mode
given by $\delta s\propto s$, but this is eliminated from the integration
because it is not consistent with the area 
constraint. Now there are two constraints
on the $\delta s$, so that two factors are missing in the numerator. One has
instead a factor $\sqrt{2\pi\sum\bar s}$ in the numerator of (\ref{sqrt}).
It is easy to see that each $\bar s$ is proportional to $A$, so that
each of these factors is proportional to $\sqrt A$ and overall
there is a factor of $\frac{1}{\sqrt A}$.
The number of states with spin projection zero is thus
\bb
N_0={1\over \sqrt{A}}\exp ({\lambda A\over 4\pi\gamma\ell_P^2}),
\ee
where constant factors have been ignored
and $\lambda$ is determined by the condition \cite{bhe1,bhe2}
\bb
\sum_{jm} e^{-2\lambda\sqrt{j(j+1)}}=1,\label{gammac}
\ee
which follows from (\ref{cond}).

Finally we have to take into account the possibility of $\sum ms$ being
different from zero but equal to zero modulo $n$.
If instead of zero, the spin projection is to be $M$, say, the number
of states will change from $N_0$. Now it is first
necessary to restore $\alpha$ which will not vanish any more.
In this case, the Lagrange multipliers $\lambda, \alpha$ satisfy the
equation (\ref{cond}). This cannot determine both parameters, but can be
solved in principle for $\lambda(\alpha)$. Note that $\bar s$ now depends
on $\alpha$ and the exponential factor in the number of configurations
changes to
\bb
\exp({\lambda(\alpha) A\over 4\pi\gamma\ell_P^2} +\alpha M)
\ee
The projection constraint (\ref{spinp}) now takes the form 
\bb
\sum me^{-2\lambda\sqrt{j(j+1)}-\alpha m}=\frac{M}{\sum\bar s}.
\ee
So although $\alpha$ is different from zero, in the case $M\ll A$, it is 
small. It follows from (\ref{cond}) that
\bb
\lambda'(0)=0
\ee
and
\bb
{M\over A/(4\pi\gamma\ell_P^2)}=-\lambda'(\alpha)=
-\alpha\lambda''(0).
\ee
Furthermore,
\bb
{\lambda(\alpha) A\over 4\pi\gamma\ell_P^2} +\alpha M&=&
(\lambda(\alpha)-\alpha^2\lambda''(0)){A\over 4\pi\gamma\ell_P^2}\nonumber\\
&=& (\lambda(0)-{\alpha^2\over 2}\lambda''(0)) {A\over 4\pi\gamma\ell_P^2}\nonumber\\
&=&\lambda(0){A\over 4\pi\gamma\ell_P^2}-
\frac{M^2}{2\lambda''(0)}
{4\pi\gamma\ell_P^2\over A}
\ee
Note that $\lambda(0)$ is the same as what was called $\lambda$ when the
constraint $\sum ms=0$ was imposed and is determined by the constraint (\ref{gammac}). 
Since from (\ref{cond})
\bb
\lambda''(0)=\frac{\sum m^2e^{-2\lambda(0)\sqrt{j(j+1)}}}{2\sum\sqrt{j(j+1)}e^{-2\lambda(0)\sqrt{j(j+1)}}},
\ee
which is positive, independent of $A,M$ and $\sim o(1)$, 
we can write \cite{bhe1, mirik}
\bb
N_M= N_0e^{-2\pi\gamma\ell_P^2M^2/(\lambda''(0)A)}.
\ee
Since $M=0$ mod $n$, we have to sum $N_M$ over the values $rn$, where 
$r=0,\pm 1, \pm2,...,$ and there arises a factor
\bb
\sum e^{-2\pi\gamma\ell_P^2r^2n^2/(\lambda''(0)A)}
\ee
which, on approximation by an integral over $r$, is seen to involve a factor
$\sqrt{A}/n$, cancelling the square root in $N_0$. We find finally
\bb
N= \frac{1}{n}\exp ({\lambda A\over 4\pi\gamma\ell_P^2}),
\ee
implying that the entropy has no logarithmic correction in $A$.
The $1/\sqrt A$ reduction arose on use of the strict projection constraint.
The dilution of the constraint, which increases the number of states,
removes that reduction factor and introduces a smaller reduction $1/n$. 

It should be pointed out that it has been argued \cite{agap} that
$\lambda(0)=\pi\gamma$ (in this case it will fix $\gamma$) 
and the entropy is exactly a quarter of $A$ in Planck units if $\log n$
is ignored.

\section*{Conclusion}

The main difference between the calculations done here or in \cite{pm} with
earlier loop quantum gravity calculations is the use of small $k$, {\em i.e.},
small Chern-Simons level compared to the area eigenvalue. The level is 
determined by the
classical horizon area, which is to be distinguished from the area eigenvalue
$A$ arising upon quantization. In the earlier literature, the
classical area or Chern-Simons level has been taken to be large,
which amounts to considering the area eigenvalue 
as restricted to be less than the classical area and that
led to logarithmic correction terms. 
The present calculations show that if one considers
area eigenvalues much larger than the fixed Chern-Simons level or
classical area, 
the degeneracy is still exponential in the area eigenvalue
but it is a pure exponential and
the entropy ceases to have a logarithmic correction term involving the area
eigenvalue. There is still a logarithmic correction, however, and it involves
the Chern-Simons level or classical area. This did not appear in the
earlier calculations because the Chern-Simons level was taken to be large
and thereby ignored. It is when $k$ is finite that the
correction involves this quantity rather than the area eigenvalue.
The correction appears to involve the smaller of the two quantities --
the area eigenvalue or the classical area.

The appearance of $k$ in the expression for the entropy is new and may cause
some surprise. However, the states are defined by the projection constraint,
which explicitly involves this parameter. This is the technical reason for the
$k$-dependence of the degeneracy. This is analogous to the familiar dependence
of energy eigenstates of the particle in a box on the classical length $L$,
which enters the problem through the boundary condition on the eigenfunctions.
The area $A$, like the energy, is a physical observable, but neither $k$ nor $L$
is an observable in the same sense as $A$ or the energy. They enter the problem
as parameters and have no quantum interpretation
because they refer to the classical area or the classical size of the box
which are not directly quantized in the standard treatment of the problem.
The parameters $k$ or $L$ determine the detailed spectrum of states from which 
they can also be determined.

A $k$ dependence is also observed \cite{pm} in the SU(2) formulation with finite
$k$ \cite{k} where there is no projection constraint, but the number of
states is explicitly known from the SU$_q$(2) theory to involve $k$.
If the numerical calculations of \cite{cor} are repeated with the full
projection constraint involving a finite $k$, the logarithmic terms
for large and small area eigenvalues can be checked.

\end{document}